\documentclass[aps,pra,twocolumn,notitlepage,superscriptaddress,showkeys,longbibliography]{revtex4-2}
\date{\today}
\usepackage{amsmath,amssymb}
\usepackage{epsfig}
\usepackage{epstopdf}
\usepackage{graphicx}
\usepackage{dcolumn}
\usepackage{bm}
\usepackage[colorlinks,linkcolor=blue,hyperindex,CJKbookmarks]{hyperref}
\usepackage{float}
\usepackage{hyperref}
\usepackage{xcolor}
\usepackage[utf8]{inputenc}
\newcommand{\beq}{\begin{equation}}
\newcommand{\eeq}{\end{equation}}
\newcommand{\bea}{\begin{eqnarray}}
\newcommand{\eea}{\end{eqnarray}}

\begin{document}
\title{Electronic structure and hinge states of strained half-Heusler compounds LiSbZn and LiBiZn}

\author{Sanjib Kumar Das}
\affiliation{IFW Dresden and W{\"u}rzburg-Dresden Cluster of Excellence ct.qmat, Helmholtzstrasse 20, 01069 Dresden, Germany}
\affiliation{Department of Physics, Lehigh University, Bethlehem, PA 18015, USA}
\affiliation{Department of Physics and Astronomy, University of Delaware, Newark, Delaware 19716, USA}

\author{Ion Cosma Fulga}
\affiliation{IFW Dresden and W{\"u}rzburg-Dresden Cluster of Excellence ct.qmat, Helmholtzstrasse 20, 01069 Dresden, Germany}

\author{Rakshanda Dhawan}
\affiliation{Department of Chemistry, Indian Institute of Technology Roorkee, Roorkee 247667, India}

\author{ Hem C. Kandpal}
\affiliation{Department of Chemistry, Indian Institute of Technology Roorkee, Roorkee 247667, India}

\author{Jeroen van den Brink}
\affiliation{Institute for Theoretical Solid State Physics, IFW Dresden, Helmholtzstrasse 20, 01069 Dresden, Germany}
\affiliation{Department of Physics, Technical University Dresden, 01062 Dresden, Germany}

\author{Jorge I. Facio}
\affiliation{Instituto Balseiro, Univ. Nacional de Cuyo, Av. Bustillo, 9500, Argentina}
\affiliation{Centro Atómico Bariloche, Instituto de Nanociencia y Nanotecnología (CNEA-CONICET), Av. Bustillo, 9500, Argentina}
\affiliation{Institute for Theoretical Solid State Physics, IFW Dresden, Helmholtzstrasse 20, 01069 Dresden, Germany}

\begin{abstract}
Half-Heusler compounds are a class of materials with great potential for the study of distinct electronic states. 
In this work, we investigate, from first-principles, the possibility of hinge modes in closely related topological phases that are tunable by moderate uni-axial strain. 
We consider two compounds: LiSbZn and LiBiZn. 
While LiSbZn has a topologically trivial band structure, the larger spin-orbit coupling of Bi causes a band inversion in LiBiZn. 
We predict the existence of topologically trivial hinge states in both cases. 
The hinge modes are affected by both the crystal termination, and the bulk topological phase transitions, albeit indirectly: 
When present, topological surface modes hybridize with the hinge states and obscure their visibility. 
Thus, we find that the most visible hinge modes occur when no band inversions are present in the material. 
Our work highlights the interplay and competition between surface and hinge modes in half-Heuslers, and may help guide the experimental search for robust boundary signatures in these materials.
\end{abstract}

\maketitle

\section{Introduction}
\label{sec:intro}

Low-dimensional boundaries of a system, such as hinges and corners, can host interesting electronic properties as recently uncovered by the discovery of higher-order topological insulators (HOTIs) and of higher-order quantized electric multipole moments~\cite{benalcazar2017quantized, Benalcazar17, PhysRevB.98.241103, PhysRevLett.119.246402, PhysRevLett.119.246401, Schindler2018, Schindler2018I, Ezawa18I, hayashi2018topological}. 
In three dimensions, a second order topological insulator is characterized by an insulating bulk, insulating surfaces and metallic hinges. 
Elementary Bi has been found to offer a material realization of such phase when crystals are cut preserving the bulk trigonal symmetry \cite{Schindler2018, Nayak2019}. 
Classifications of symmetry-protected HOTI phases have been performed \cite{PhysRevLett.123.196401, PhysRevLett.123.266802, geier2020bulk, simon2021higher, Sanjib23} and the study of hinge states in a plethora of platforms is nowadays very active \cite{Franca18, PhysRevB.103.115428, PhysRevB.103.184510, zhao2020topological, fang2021dirac}.

The possibility of hinge states in gapless semimetals has also been recognized \cite{PhysRevLett.125.266804, PhysRevLett.125.146401, Wieder2020obs, Fang20} and the quest for specific materials where to study them is naturally important.
Reference \cite{Wieder2020obs} showed the existence of hinge states in models relevant for various candidate Dirac semimetals, such as Cd$_3$As$_2$ and KMgBi.
Reference \cite{Wang19} argued about the importance of hinge states in $\beta$-MoTe$_2$ for the interpretation of its surface electronic structure while Ref.~\cite{choi2020evidence} reported evidence of hinge states in the related compound WTe$_2$.

Recently, it has been suggested the possibility of HOTI phases in the broad set of half-Heusler (hH) compounds \cite{PhysRevB.101.121301}.
Half-Heuslers are ternary compounds with the space group (SG) No.~216 \cite{chadov2010tunable, lin2010half, PhysRevLett.105.096404}. 
They present at low energy a very similar electronic structure to that of CdTe or HgTe, the latter one a reference example of nontrivial electronic topology since the first predictions \cite{Bernevig2006, PhysRevB.76.045302} and experimental confirmation \cite{Konig766, PhysRevLett.106.126803} of hosting a topological insulating phase under strain.
The richness of HgTe goes hand in hand with the simplicity of its band structure, the hallmark of which is an inversion between the so-called $\Gamma_6$ and $\Gamma_8$ bands \cite{Khaetskii}. 
In the cubic phase, the $\Gamma_8$ bands exhibit a fourfold crossing at low energies which enforces a semimetallic phase.
An appropriate reduction of the symmetry opens a gap, yielding a strong topological insulator (TI).
 
The additional atom in the unit cell of hHs can serve different purposes, such as providing a knob to change the unit-cell volume, which in turn reflects on the electronic structure, or stabilizing various long-range ordered phases \cite{graf2011simple}, making hHs an interesting platform to explore the interplay between such diversity and topological properties of the electronic structure.
This potential naturally motivated extensive \textit{ab initio} computational searches \cite{PhysRevB.82.125208, PhysRevB.86.075316, PhysRevB.91.094107} that yielded as a result the identification of dozens of band-inverted Hhs, namely compounds in which the bands follow the hierarchy $\varepsilon_{\Gamma_8} > \varepsilon_{\Gamma_6}$, 
the surface states of some of which were experimentally studied \cite{logan2016observation, liu2016observation, PhysRevB.108.165154}. 
Later, the possibility of realizing the Weyl semimetal phase in different members of the class further enriched the prospects of non-trivial topology in hHs. 
This was shown in studies that considered different mechanisms that act to reduce the cubic symmetry, including magnetic fields \cite{PhysRevB.95.161306, suzuki2016large, hirschberger2016chiral} or uniaxial strain \cite{ruan2016symmetry}.
These axial perturbations break the three-fold rotational symmetries that protect the fourfold crossing of the $\Gamma_8$ bands \cite{Bradlyn2016}, opening a gap in these bands and eventually leading to the creation of Weyl nodes.

The above-mentioned suggestion of HOTI phases in hHs is based on a study of a four-band model reasonably accurate to describe the $\Gamma_8$ bands \cite{PhysRevB.101.121301}. 
In this work, we investigate the problem by means of density-functional theory (DFT) calculations and of ab-initio derived tight-binding models. 
Our calculations show the existence of electronic states on certain hinges of hHs with either $\varepsilon_{\Gamma_8} > \varepsilon_{\Gamma_6}$ or with $\varepsilon_{\Gamma_8} < \varepsilon_{\Gamma_6}$. 
Importantly, we find these hinge states to be topologically trivial. 
Further, we show that they are critically affected by the topology of the bulk electronic structure due to trivial hybridization effects: 
Topological surface states, when present, hybridize with the hinge modes which, as a consequence, avoid being exponentially localized around the hinges.

This paper is organized as follows. In Sec.~\ref{sec:methods}, we describe methodological aspects of this work. 
In Sec.~\ref{sec:LihHs}, we present the bulk and surface electronic structure of the lithium-based half-Heusler compounds LiBiZn and LiSbZn. 
In Sec.~\ref{sec:loweffH}, we discuss relevant Hamiltonians that describe the low energy electronic properties of these compounds  
and in Sec.~\ref{sec:hinge} we present hinge band structure calculations.  Sec.~\ref{sec:conc} contains our concluding remarks.    
The Appendix.~\ref{app:TB} provides 8-band tight-binding Hamiltonians suitable to describe LiBiZn and LiSbZn.
Data and code that reproduce our main results are available on Zenodo \cite{Zenodo_code}.

\section{Methods}
\label{sec:methods}

Density functional theory (DFT) calculations were performed with the FPLO code \cite{PhysRevB.59.1743} version 18.57, treating the spin-orbit coupling in the fully-relativistic four-component formalism and performing numerical Brillouin zone integrations with the tetrahedron method using a mesh of $30\times30\times30$ subdivisions.
We relaxed the crystal structure by minimization of the total energy. 
For the simulation of uniaxial strain, we fixed the deformation of the lattice parameter $c$, and relaxed the perpendicular lattice parameters under the constraint $a=b$.
To parametrize the deformation, we define $\delta=(c_0-c)/c_0$, where $c_0$ is the equilibrium lattice parameter.
Our primary goal is to identify which phases are accessible under moderate uniaxial strain, specifically, whether a HOTI phase should be expected. 
For clarity in the band-structure analysis, we use representative strains of $\delta=\pm2\%$. 
Although such values may be challenging to realize in this materials family, they sharpen the features central to our discussion. 
Importantly, our qualitative conclusions about the presence or absence of a HOTI phase do not depend on the precise magnitude of $\delta$.

Starting from the DFT results, tight-binding models were obtained by construction of Wannier functions with the projective method as implemented in FPLO~\cite{koepernik23}.
Based on the analysis of the orbital-projected density of states, 
we consider an eight band model that includes Bi-6p (or Sb-5p) and Zn-4s orbitals. 
These tight-binding models were used for the study of the electronic structure at surfaces and hinges.
For the former, we considered semi-infinite slabs using the PYFPLO module of the FPLO code, while for the latter we solved finite systems using own codes based on Kwant \cite{Groth2014} (more details are presented in Section \ref{sec:hinge}).

\section{Lithium half-Heuslers}
\label{sec:LihHs}

\begin{figure}[t!]
\centering
\includegraphics[width=0.9\columnwidth,angle=0]{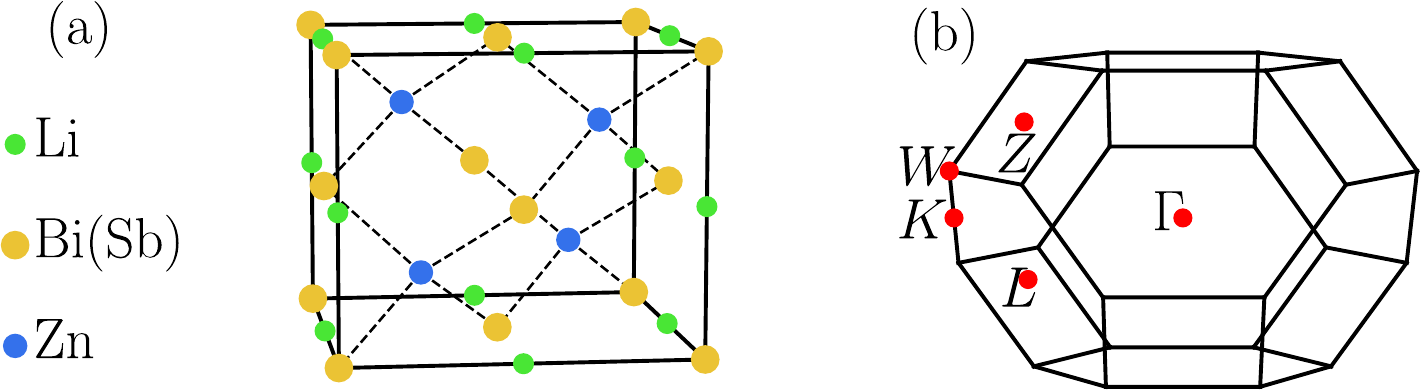}
\vspace{0.2cm}
\includegraphics[width=\columnwidth,angle=0]{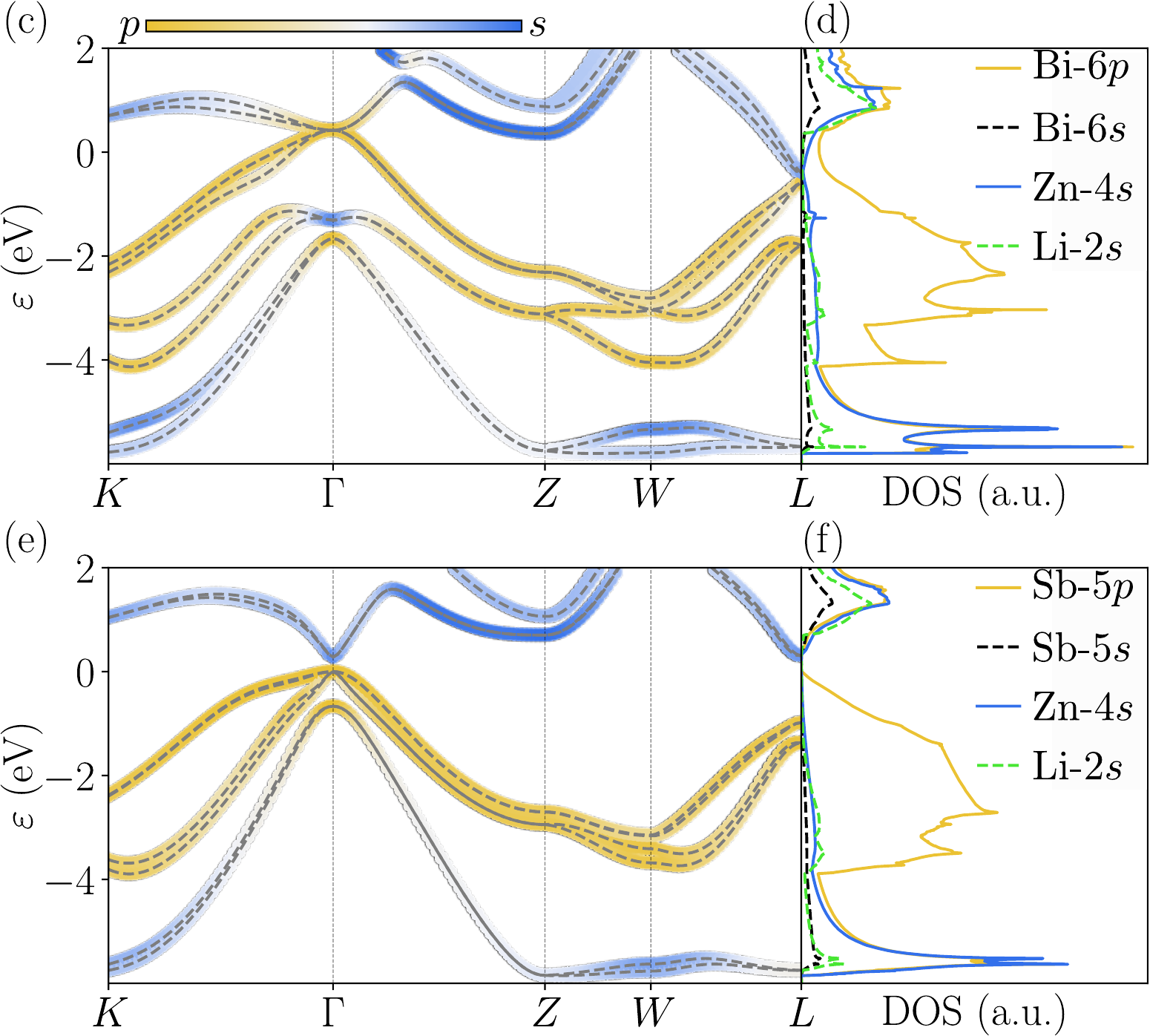}
\caption{
(a) Crystal structure. 
(b) Brillouin zone.
(c,d) Band structure and density of states of LiBiZn and LiSbZn, respectively.
The weight associated with Bi-6p (or Sb-5p) is depicted in orange while that of Zn-4s is shown in blue.
}
\label{fig:bands_cubic} 
\end{figure}

The crystal structure of a hH compound of formula $XYZ$ can be regarded as two displaced zinc-blende lattices formed by $XZ$ and $YZ$ pairs of atoms [Fig.~\ref{fig:bands_cubic}(a)].
With this notation, $X$ and $Y$ occupy the crystallographically equivalent sites $4a$ and $4b$, forming a  rock salt structure, while $Z$ is positioned at the site $4c$.
We will consider the cases $X$$=$Li, $Y$$=$Bi or Sb and $Z$$=$Zn.

We choose Li-based hHs because, as we will show, their simplicity facilitate the construction of low-energy tight-binding models that accurately reproduce the density-functional calculations with a minimum effort in terms of number of bands. 
This allows us to tackle the problem of studying the hinge electronic structure using realistic tight-binding Hamiltonians. 

While both LiSbZn and LiBiZn have originally been studied in the hexagonal phase \cite{PhysRevB.96.115203,Gao21}, the former has also been grown in the cubic phase \cite{white2016polytypism} and the latter has recently been predicted to present a cubic phase as ground state \cite{chopra2019first}.
We notice that within the same cubic space group, there are two variants defined by the crystallographic site occupied by the Zn atoms. 
In particular, in the family LiZnT (T$=$N, P, As), the Zn atoms are found in the $Y$ position. 
On the other hand, Ref.~\cite{white2016polytypism} has shown that the compound LiSbZn is different, since Zn occupies the $Z$ position [see Fig.~\ref{fig:bands_cubic}(a)]. 
We have compared the total energy of these two variants for LiBiZn and LiSbZn, obtaining that the structure with Zn in the $Z$ site has lower energy. 
We notice that for both structures the Bi and Zn atoms --- which provide the dominant contribution near the Fermi energy, as will be shown below --- form a zinc-blende lattice and therefore, both variants have similar low-energy band structures.

We obtain for the fully relaxed compounds the lattice parameters $a=6.584\,$\AA\,\, for LiBiZn and $a=6.350\,$\AA\,\, for LiSbZn.
Figure \ref{fig:bands_cubic} (c, d) shows the band structures and the orbital-projected density of states $D(\varepsilon)$ of LiBiZn and of LiSbZn, respectively.  
In both cases, Bi-$6p$ or Sb-$5p$ states dominate the electronic structure in the energy windows shown and exhibit a substantial hybridization with Zn-$4s$ and Li-$2s$.
The main difference between compounds is the relative position of the $\Gamma_6$ and $\Gamma_8$ bands, which accommodate in total four electrons per unit cell. 
In LiSbZn, the energies of these bands obey $\varepsilon_{\Gamma_8} < \varepsilon_{\Gamma_6}$, so that the $\Gamma_8$ bands are completely filled and the compound is an insulator. 
In LiBiZn, the opposite arrangement leads to half-filled $\Gamma_8$ bands, enforcing a semimetallic state. 
In this case, a clear $s$-$p$ inversion is noticeable near $\Gamma$.
In Appendix \ref{app:TB} we present an eight-band tight-binding model suitable to describe both compounds. 
An analysis of the model parameters indicates that the main difference between compounds is the larger local spin-orbit coupling in the Bi-based compound.
Thus, these compounds having $\varepsilon_{\Gamma_8} > \varepsilon_{\Gamma_6}$ (LiBiZn) or $\varepsilon_{\Gamma_8} < \varepsilon_{\Gamma_6}$ (LiSbZn) provide a framework for the study of the hinge electronic structure of both band-inverted and normal systems having very similar chemical and physical properties (e.g. lattice parameters and orbital composition of low-energy bands).

\begin{figure}[t!]
\centering
\includegraphics[width=0.5\textwidth,angle=0]{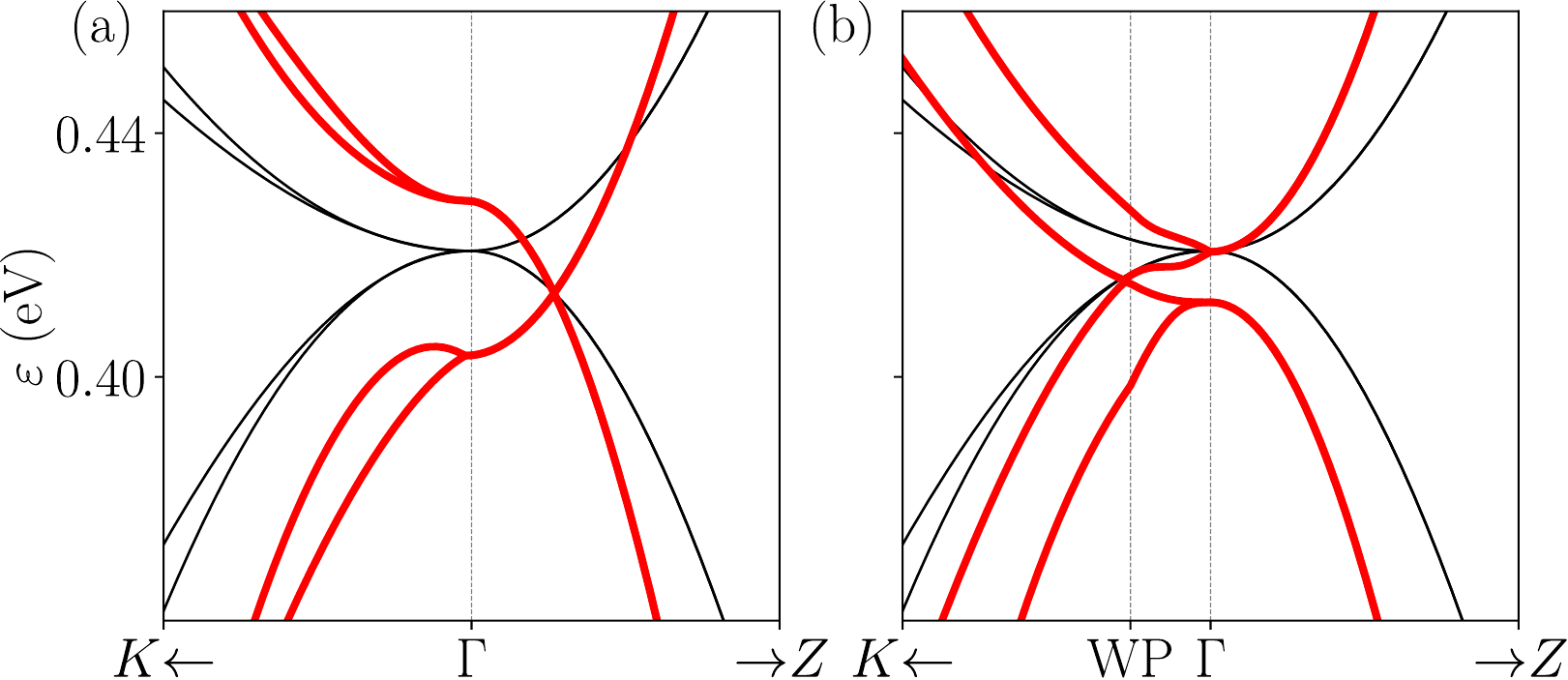}
\caption{
(a) Band structure of LiBiZn under uniaxial strain $\delta=-0.02$. 
Thinner black curves are the bands without strain. 
(b) Band structure for $\delta=0.02$. 
The path chosen includes one of the eight Weyl nodes found in the $k_z=0$ plane, of coordinates $\frac{2\pi}{a}(0.017,0.006,0)$.
}
\label{fig:bands_uniaxial} 
\end{figure}

The fourfold crossing present at $\Gamma_8$ is protected by the threefold rotation symmetries along the unit-cell diagonals \cite{Bradlyn2016}.
Therefore, perturbations that reduce the cubic symmetry will generically gap out this crossing.
Here we consider the case of uniaxial strain along the $\hat{z}$ direction, either compressing (negative $\delta$ in our convention) or stretching the system.
Under such perturbations, the crystal symmetry becomes I$\bar{4}$m2 (SG 119).
Figures \ref{fig:bands_uniaxial} (a) and (b) show the band structures of LiBiZn near $\Gamma$ corresponding to $\delta=0,\pm 0.02$. 
For both signs of $\delta$, we find that the $\Gamma_8$ bands remain gapless in a wide range of strain $-0.04<\delta < 0.04$.
For $\delta<0$, a Dirac cone along $\Gamma Z$ closes the gap.
For $\delta>0$, Weyl nodes closing the gap are found in the $k_z=0$ plane. 
Similarly, in LiSbZn strain opens a gap at the fourfold crossing. In this case, the system remains insulating with only a weak change in fundamental gap formed between the $\Gamma_8$ and $\Gamma_6$ bands (not shown).

Lastly, we notice the  existence of planes defined by three of the time-reversal invariant momenta (more details in Section \ref{sec:loweffH}) which present for finite strain a well defined  gap in the full Brillouin zone for any sign of strain and have a nontrivial $\mathbb{Z}_2$ index. 
Thus, for finite $\delta$ the electronic structure is characterized by either Weyl or Dirac points which close the gap of the $\Gamma_8$ bands, but which coexist with two-dimensional planes that realize quantum spin Hall insulating (QSHI) phases.

\begin{figure}[t!]
\centering
\includegraphics[width=0.5\textwidth,angle=0]{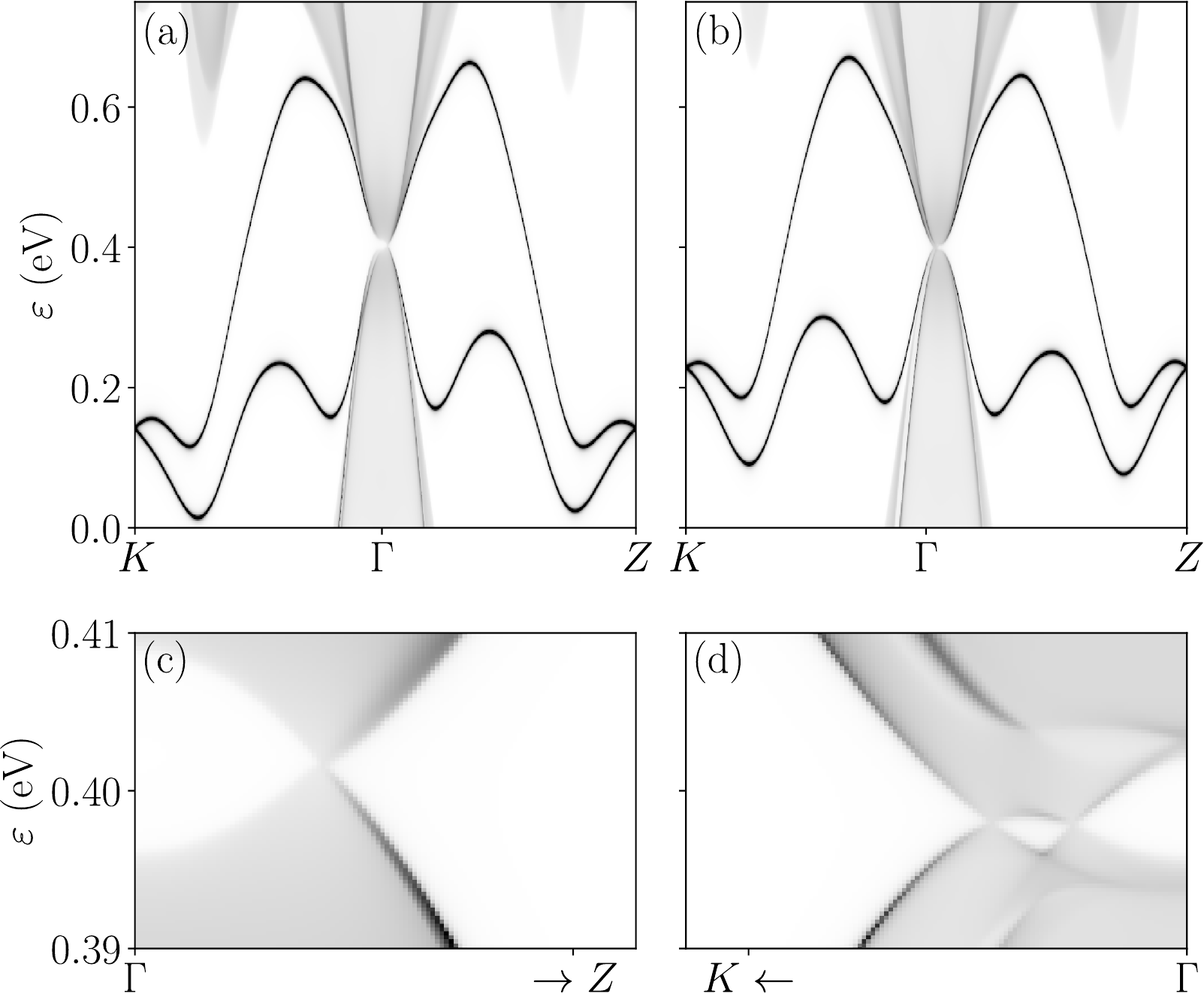}
\caption{
(a) (1$\bar{1}$0) Zn-terminated surface spectral density of LiBiZn under uniaxial strain $\delta=-0.02$. 
The panel on the right is a zoom of the white square in the left panel. 
(b) Analogous results for strain $\delta=0.02$. 
(c), (d) Zoom of the data in (a), (b), respectively. 
The path width in the zooms is 0.04$\times 2\pi/a$.
}
\label{fig:bands_surface_uniaxial} 
\end{figure}

To conclude this Section, we briefly present how these properties influence the surface spectral features.
We focus on LiBiZn and consider the Zn-terminated (1$\bar{1}$0) surface, which is perpendicular to the direction in which strain is applied.
Figures \ref{fig:bands_surface_uniaxial} (a) and (b) show results for $\delta=-0.02$ and $\delta=0.02$, respectively.
The path chosen includes segments perpendicular ($K \Gamma$) or parallel ($\Gamma Z$) to  the direction of the strain.
Surface states with Dirac cones located at the Brillouin zone border ($K$ and $Z$) are observed for both signs of strain.
We notice that for a LiBi-terminated surface, while the connectivity provided by the surface states between the projection of the bulk valence and conduction bands is the same, the Dirac cone at $Z$ or $K$ is pushed into the bulk projection (not shown).
As shown in Figures \ref{fig:bands_surface_uniaxial} (c) and (d), the main role of strain is observed near $\Gamma$: the spectrum becomes gapped at $\Gamma$ and gapless at the projection of the Dirac cone (along $\Gamma Z$ and $\delta < 0$) or of the Weyl nodes (along $K \Gamma$ and $\delta > 0$).

\section{Low-energy models}
\label{sec:loweffH}

The salient feature of the low energy band structure of band-inverted hHs is the seemingly quadratic touching at $\Gamma$, which involves four bands.
A model that describes a quadratic fermionic band touching at an isolated point and is often invoked as relevant for hHs materials is the spin 3/2 Luttinger Hamiltonian. 
This model is rooted in the $\mathbf{k} \cdot \mathbf{p}$ perturbation theory \cite{PhysRev.102.1030} and has been used for investigating, e.g., spin-Hall conductivity \cite{PhysRevB.69.235206}, non-Fermi-liquid topological phases \cite{PhysRevLett.111.206401}, light-matter interaction in Weyl semimetals \cite{PhysRevB.97.205402} and also superconducting pairing properties \cite{PhysRevB.96.214514,PhysRevB.99.054505}. 
More recently, the topological properties of this model have been analyzed in Ref.~\cite{PhysRevB.101.121301}, where it was found that uniaxial strain leads to the formation of HOTI phases.
Thus, it is naturally important to understand how these results apply to hHs.

Following Ref.~\cite{PhysRevB.101.121301}, we consider the model Hamiltonian
\begin{equation} \label{eq:Lutt}
\begin{split}
\mathcal{H}(\mathbf{k}) & = t \sum \limits_{j=1}^{2} \sin(k_{j}a) \gamma_{j} \\
& + \Big[ -t_{3}\cos(k_{3}a)+m_{z}+t_{0}\sum \limits_{j=1}^{2}
[1-\cos(k_{j}a)] \Big] \gamma_{3}\\
& + \Delta_{1}[\cos(k_{2}a)-\cos(k_{1}a)] \gamma_{4} + \Delta_{2} \sin(k_{1}a) \sin(k_{2}a) \gamma_{5},
\end{split}
\end{equation}
where $a$ is the lattice constant, $\mathbf{k}=(k_1,k_2,k_3)$ is the momentum vector, and $\gamma_{j}$, $1 \leq j \leq 5$, are the mutually anticommuting $4 \times 4$ Hermitian matrices satisfying the algebra $\lbrace{\gamma_{i},{\gamma_{j}}\rbrace = 2 \delta_{ij}}$. 
These can be represented as $\gamma_{1} = \sigma_{3} \tau_{1}$, $\gamma_{2} = \sigma_{0} \tau_{2}$, $\gamma_{3} = \sigma_{0} \tau_{3}$, $\gamma_{4} = \sigma_{1} \tau_{1}$, $\gamma_{5} = \sigma_{2} \tau_{1}$, where $\sigma$ and $\tau$ Pauli matrices encode the spin and orbital degrees of freedom, respectively.
As shown by Ref.~\cite{PhysRevB.101.121301}, expansion of this model around $\Gamma$ can be mapped to the Luttinger Hamiltonian.
The model describes well the low-energy dispersion of band-inverted hHs, up to differences stemming from distinct symmetries. 
In particular, for real coefficients, $\mathcal{H}(\mathbf{k})$ is invariant at each $\mathbf{k}$ under the antiunitary operator $\tilde{T} = \sigma_2 \tau_3 K$  (not the physical time-reversal), where $K$ denotes complex conjugation and $\tilde{T}^2=-1$. 
Consequently, at generic $\mathbf{k}$ the bands are doubly degenerate. 
In actual hHs, spatial inversion is broken and spin-orbit coupling lifts the spin degeneracy, with doubly degenerate bands persisting only along high-symmetry directions (e.g. $\Gamma$-$X$).

The term $\delta=(-t_3 \cos k_{3}a+m_z) \gamma_3$ mimics the effects of strain along the $[001]$ direction \footnote{For consistency with our DFT results, we define $\delta$ with an opposite sign to Ref.~\cite{PhysRevB.101.121301}.}. 
Depending on the sign of $\delta$ at $k_3=0$, this term leads the model Eq.~\eqref{eq:Lutt} to either a Dirac semimetal phase ($t_3>m_z$), or to a stack of quantum spin Hall insulator (3D) phases ($t_3<m_z$). 
Interestingly, these phases support higher-order topological modes along certain hinges \cite{PhysRevB.101.121301}. 

We observe that the topological properties of the Hamiltonian Eq.~\eqref{eq:Lutt} are in general different from those of hHs we consider.
First, in the case of LiBiZn, the stack of QSHI phases is instead replaced by a Weyl semimetal phase, as shown in Fig.~\ref{fig:bands_surface_uniaxial}(b).
Second, and more fundamental, as shown in earlier studies in HgTe \cite{PhysRevB.77.125319} and also in some hHs \cite{PhysRevB.95.235158,PhysRevMaterials.8.124201}, to faithfully describe not only the energy dispersion in a given energy window but also global topological properties of the full (occupied) band structure, a model for hHs must have at least six orbitals per primitive unit cell \cite{PhysRevB.95.235158}.
This follows from the fact that the band inversion that drives the non-trivial topology is between the quartet of bands $\Gamma_8$ and the doublet $\Gamma_6$. 
The difference between the effects of strain on the model Eq.~\eqref{eq:Lutt} and on a model for hHs containing six (or more) bands can be appreciated noticing that, as mentioned above for LiBiZn, band-inverted hHs have in the Brillouin zone planes which realize a QSHI phase under finite strain of \textit{any} sign. 
On the contrary, in the case of the model Eq.~\eqref{eq:Lutt}, these planes undergo a phase transition from a QSHI to a trivial insulator as the strain changes sign.

\begin{figure}[tb]
\centering
\includegraphics[width=\columnwidth,angle=0]{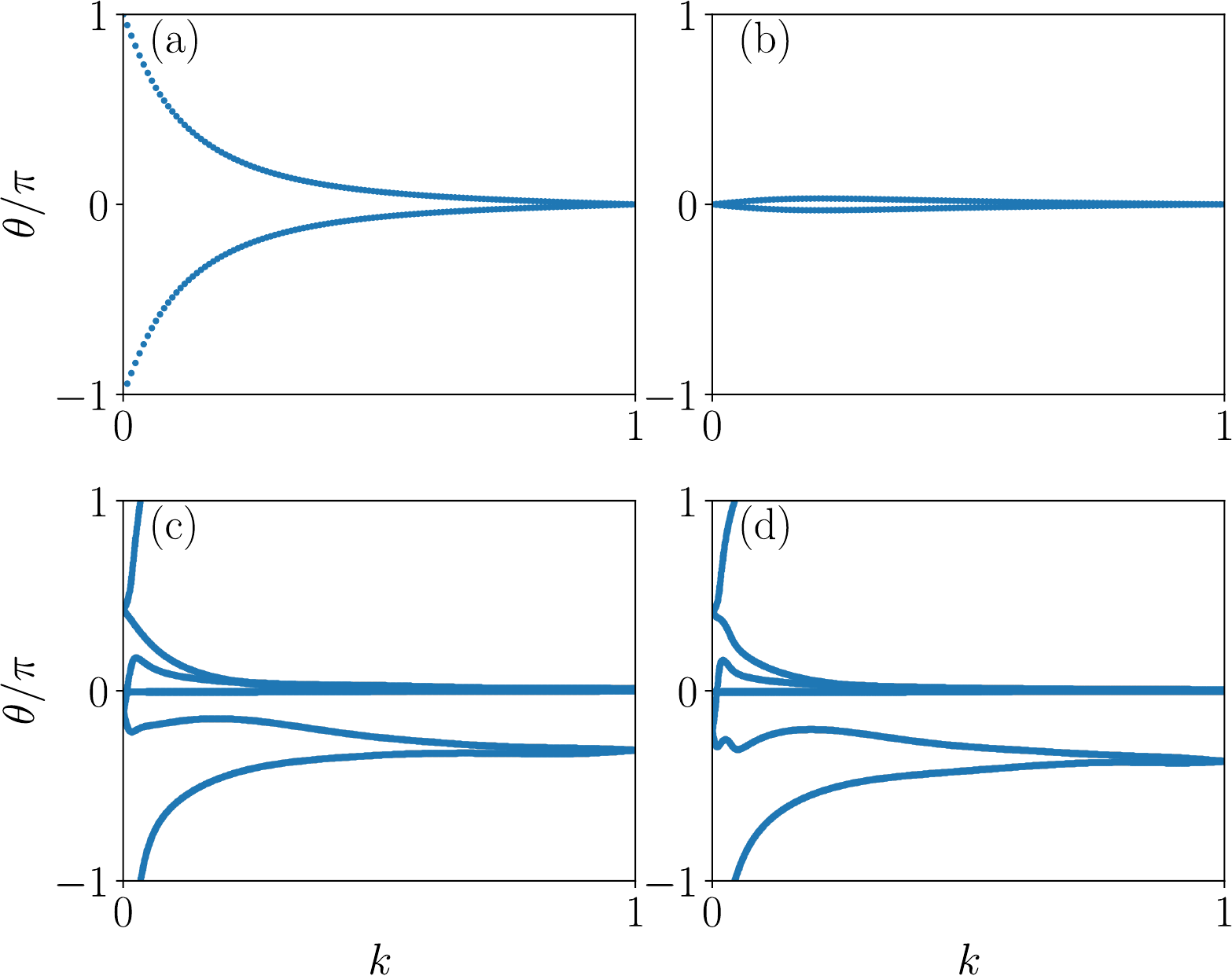}
\caption{
(a) Wilson loop in the $k_z=0$ plane for the four band model Eq.~\eqref{eq:Lutt} with $\delta<0$. 
(b) Same for $\delta>0$.
(c) Wannier center evolution in the plane defined by $\Gamma$ and two adjacent L points for LiZnBi under compressive strain.
(d) Same as (c) in the case of tensile strain. 
}
\label{fig:wloops} 
\end{figure}

To illustrate the difference in topology, Figs.~\ref{fig:wloops}(a) and (b) show the Wilson loop computed in the $k_3=0$ plane for positive or negative strain, respectively, for the model Eq.~\eqref{eq:Lutt}. 
The different winding indicates a change in the value of the $\mathbb{Z}_2$ invariant for different signs of strain.
On the other hand, the $\mathbb{Z}_2$ invariant in LiBiZn remains nontrivial independently of the sign of strain, as indicated by the evolution of the Wannier centers shown in Figs.~\ref{fig:wloops}(c) and (d) for compressive and tensile strain, respectively. 
Therefore, the electronic structure of both surfaces and hinges can be very different between a system described by the model Eq.~\eqref{eq:Lutt} and hHs.
Particularly, in band-inverted hHs the first-order topology makes the surfaces gapless, so that these systems do not qualify as HOTIs.

\begin{figure*}[t]
\centering
\includegraphics[width=\textwidth]{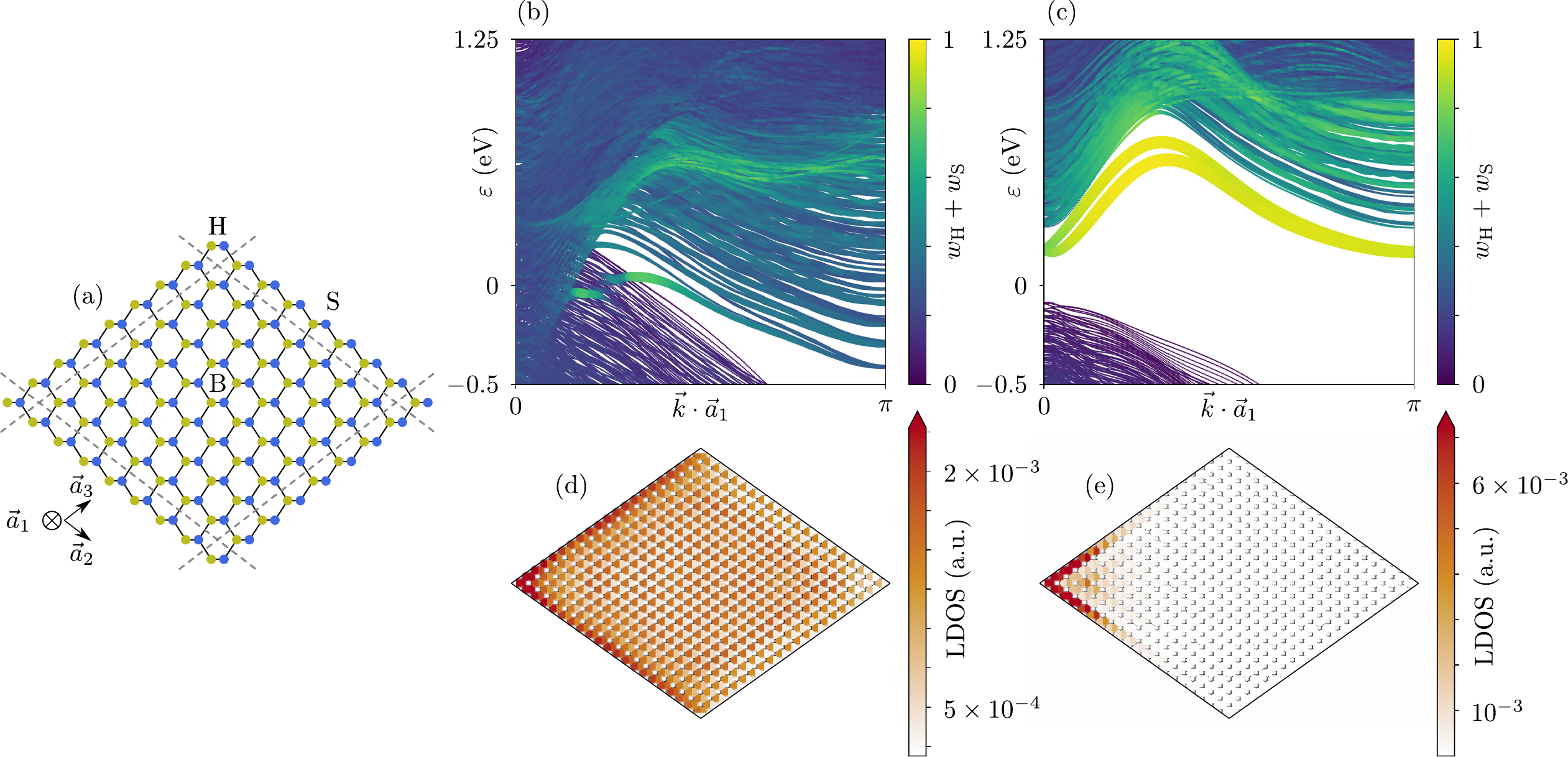}
\caption{
Panel (a): Cross-section of the prism geometry, infinite along $\vec{a}_1$ and finite along $\vec{a}_{2,3}$. 
Yellow circles correspond to Bi (or Sb) atoms and blue circles correspond to Zn. 
Bulk, surface, and hinge regions are labelled as B, S, and H, respectively. 
Notice also that the downfolded Hamiltonian does not contain orbitals at the Li site. 
Panels (b, c): Band structure in the absence of strain in the case of LiBiZn (b) and LiSbZn (c). 
In these simulations, the system size is 20 unit cells along both $\vec{a}_2$ and $\vec{a}_3$.
Lighter colors denote states localized at the boundaries of the prism, whereas thicker lines denote states localized at the hinges, as explained in the text.
Panels (d, e): The local density of states of the infinite prism, averaged over momentum, is shown for LiBiZn at $E=0$ (d), and for LiSbZn at $E=0.2$ eV (e).
}
\label{fig:rhombus} 
\end{figure*}

\section{Hinge electronic structure}
\label{sec:hinge}

In this section, we present hinge band structure calculations for LiBiZn and LiSbZn.
These are based on tight-binding Hamiltonians associated with Wannier functions constructed from DFT calculations.
As explained above, a six-band model would be the minimal option to correctly capture the topology of band-inverted hHs. 
However, it is problematic to obtain such a model from a Wannierization process consisting in the projection of the Kohn-Sham wave functions onto pre-defined local orbitals, essentially because the low-energy six bands have a strong mixture of Bi-6p and s-like orbitals (primarily from Zn but also in smaller amount from Li and Bi as well). 
This makes the definition of the local orbitals onto which project unknown. 
Therefore, at the expense of a numerically more demanding hinge bandstructure calculation, we shall work with 8-band models including all Bi 6p (or Sb 5p) and Zn 4s orbitals, as explained in Section \ref{sec:methods}.

The lattice Bravais vectors $\vec{a}_1,\vec{a}_2,\vec{a}_3$ read
\begin{equation}
\begin{pmatrix}
\vec{a}_1 \\
\vec{a}_2 \\
\vec{a}_3
\end{pmatrix} =  \frac{a}{2}
\begin{pmatrix}
0 & 1 & 1\\
1 & 0 & 1\\
1 & 1 & 0
\end{pmatrix} \begin{pmatrix}
\hat{x} \\
\hat{y} \\
\hat{z}
\end{pmatrix},
\end{equation}
where $\hat{x}$, $\hat{y}$, and $\hat{z}$ are unit vectors in the three Cartesian directions.
Figure \ref{fig:rhombus}(a) illustrates the system we consider: a prism which is infinite along $\vec{a}_1$ and finite along $\vec{a}_{2,3}$. 
Two of the surfaces are terminated at Bi/Sb sites and the remaining two at Zn. 
There are three inequivalent hinges: one results from the intersection of Zn planes, one from Bi (or Sb) planes and two from the intersection of a Bi (or Sb) plane with a Zn plane.
The calculations were performed for a system of size $20 \times 20$.
To study the real-space distribution of the wavefunctions, we divide the system in regions associated with the hinges (four regions labelled H), the surfaces (four regions labelled S) and the bulk (one region labelled B), as depicted in Fig.~\ref{fig:rhombus}(a). 
Accordingly, we introduce the weight functions $w_\text{H}(k_x,n)$, $w_\text{S}(k_x,n)$ and $w_\text{B}(k_x,n)$ which measure in each region the probability amplitude of the Bloch wavefunctions. These are defined as

\begin{align}\label{eq:weights}
	w_\text{R}(k_1, n) &= \sum\limits_{\mathbf{r} \in R}  |\langle \mathbf{r}|\psi_{n}(k_1, \mathbf{r}) \rangle|^2,
\end{align}    
where R labels the regions defined above, $\mathbf{r}$ a vector in the $\vec{a}_2$--$\vec{a}_3$ plane, $k_1 = \vec{k}\cdot \vec{a}_1$ is the crystal momentum, and $n$ is a band index.
 
Figures~\ref{fig:rhombus}(b) and (c) show the prism band structure in the absence of strain for LiBiZn and LiSbZn, respectively.
To highlight the electronic structure near surfaces as well as hinges, the points are colored according to $w_\text{S}+w_\text{H}$ have larger size the larger $w_\text{H}$ is.
Thus, lighter colors denote boundary states and thicker lines imply hinge states.
For LiBiZn, the gap is closed near $\Gamma$, while states with relatively large amplitudes at both the hinges and surfaces can be observed away from $\Gamma$. 
For these states, however, we obtain $w_\text{H}\sim w_\text{S}$, indicating that hinge modes are strongly hybridized with surface states. 
At similar energy, LiSbZn instead exhibits states with large amplitudes at the hinges, $w_\text{H} \gg w_\text{S}$. 

Notice that the hinge modes in LiSbZn do not interpolate between states above and below the Fermi level. 
Thus we conclude that LiSbZn is not a HOTI.
Furthermore, the energy dispersion illustrates that there is no symmetry requirement on the energy of these hinge states, such that they could in principle be shifted in energy away from the gap, producing a trivial bandstructure.

Figures~\ref{fig:rhombus}(d) and (e) present for both compounds the local density of states (LDOS) which could be measured, e.g., by scanning tunneling spectroscopy. 
In the case of LiBiZn, the LDOS is computed at $\varepsilon=0$, whereas $\varepsilon=0.2$ eV is chosen for LiSbZn.
A relatively large LDOS at the Bi/Sb hinge can be observed in both compounds. 
In line with the results above, LiSbZn  being both insulating in the bulk and at surface, the found hinge modes are exponentially localized at the hinge for all momenta and the LDOS accordingly decays very fast away from the hinge.
This can be quantified by tracking the weight function $w_{\text{H}}$ associated to the hinge states as a function of momentum. We find $w_{\text{H}}\simeq 0.36$ close to the center of the Brillouin zone, going up to values $w_{\text{H}}\simeq 0.71$ for larger $\vec{k} \cdot \vec{a}_1$ [see Fig.~\ref{fig:rhombus}(c)].
Notice that the hinge states are present only at certain corners, further indicating that their physical origin is associated with the crystal structure termination at the hinge, rather than with a bulk topological property.
On the other hand, a much larger spread of the hinge mode along the surface can be observed in LiBiZn, as indicated also by the momentum-resolved weight functions presented in Appendix \ref{app:weights}.
An analysis of the orbital composition indicates that hinge modes have a weight of $\sim 95$\% on the Bi/Sb orbitals.
Further, we find that uniaxial strain can affect the energy of the hinge modes.
Changes in the order of 10\,meV are found for a 2\% lattice deformation.

Finally, we note that the energy and the real-space probability density of these boundary-localized modes depend on the crystal termination, i.e. the shape of the prism cross-section.
We show in Appendix \ref{app:rectangle} an alternate choice of geometry, in which the cross-section is rectangular instead of rhombus-shaped. 
In this case, we observe that the modes become spread out over the entire Bi(Sb)-terminated surface.

\section{Conclusions}
\label{sec:conc}

Motivated by recent predictions of hinge modes in models for systems with biquadratic band crossings, we have analyzed based on accurate tight-binding models the surface and hinge electronic structure of two half-Heusler compounds, LiBiZn and LiSbZn.
These compounds should be representative of many others in the space group 216, the former having a band inversion and the latter being topologically trivial.
Our calculations indicate the existence of topologically trivial hinge modes in both kinds of systems. 
In LiBiZn, topological surface states can naturally hybridize with these modes, which obscures their observation. 
On the contrary, the electronic structure of the topologically trivial LiSbZn is gapped both in the bulk and at the surfaces, and the hinge modes are exponentially localized for all momenta.
Thus, our results indicate that topologically trivial members of the half-Heusler material class are prone to exhibit cleaner (free from hybridization with bulk or surface states) hinge modes than topological non-trivial compounds.

In addition, we have found that uniaxial strain reducing the cubic symmetry does not change the topology of the found hinge modes but can affect their energy. 
This might provide a controllable platform to experimentally study various problems involving electronic matter in one dimension. 
One relevant open question is the stability of the hinge modes with respect to possible surface reconstructions, known to be relevant for semiconductors within the same space group \cite{kahn1983semiconductor}. 
Experimental information of the actual hinges that naturally tend to exist in this broad material class would be of interest. 
Last, in view of their relatively small bandwidth, a second interesting question for future investigation is the fate of the found hinge modes in the presence of Coulomb interactions not considered in our tight-binding Hamiltonian description. 

\vspace{3mm}

\section{Acknowledgements}
We thank Ulrike Nitzsche for technical assistance. 
JIF would like to thank the Alexander von Humboldt Foundation for support.
This work is supported by the  Deutsche Forschungsgemeinschaft (DFG, German Research Foundation) through the W\"{u}rzburg-Dresden Cluster of Excellence on Complexity and Topology in Quantum Matter -- \emph{ct.qmat} (EXC 2147, project-id 390858490)
\bibliography{ref}

\appendix

\section{Eight-band tight-binding Hamiltonian}
\label{app:TB}
	
In this section, we provide a tight-binding model that describes the low-energy electronic structure of face-centered cubic half-Heuslers in space group 216. 
While the tight-binding Hamiltonian obtained from the Wannier functions construction naturally involves many long-range hopping parameters, we have found that a truncated Hamiltonian, which keeps only nearest and next-nearest neighbor terms provides an efficient low-energy description. 

\begin{figure}[t]
\centering
\includegraphics[width=2\columnwidth]{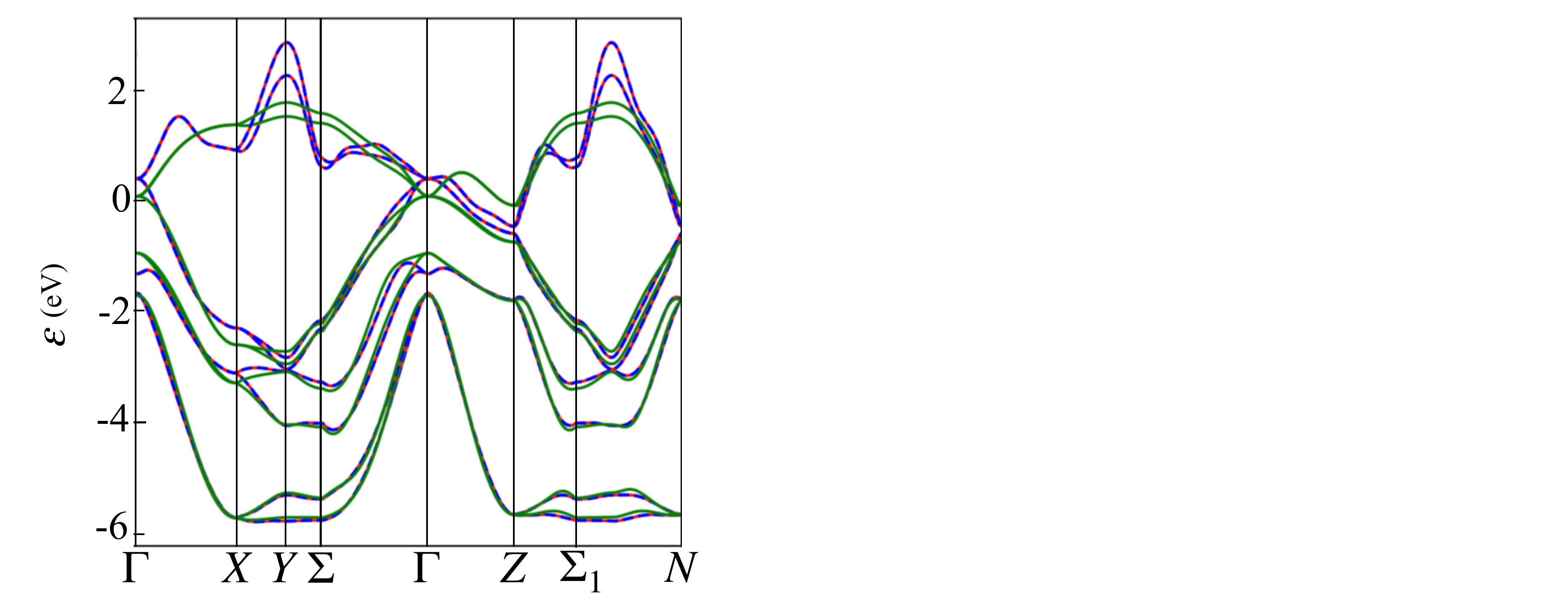}
\caption{
Band structure for LiBiZn. 
Red, blue and green colors indicate the band structures obtained from DFT, full Wannier tight-binding, and truncated Wannier tight-binding Hamiltonian, respectively. 
Qualitatively similar level of agreements between the DFT and Wannier models is obtained for LiSbZn.
}
\label{fig:bsapp} 
\end{figure}
	
Notice that the calculations in the main text are based on the tight-binding models involving all of the \textit{ab initio} obtained matrix elements.
As explained in the main text, at least six orbitals must be considered to correctly describe the topological properties of the electronic structure of hHs.
For simplicity in the DFT downfolding, we keep eight orbitals in the energy windows ($[-6,2]\,$eV). 
We use the basis  $\{|s \uparrow \rangle, |s \downarrow\rangle, |p_{y} \uparrow \rangle, |p_{y} \downarrow\rangle, |p_{z} \uparrow \rangle, |p_{z} \downarrow\rangle, |p_{x} \uparrow \rangle, |p_{x} \downarrow\rangle  \}$ where $s$ corresponds to Zn-$4s$ and $p$ to either Bi-$6p$ or Sb-$5p$.
As explained in the main text, the lattice Bravais vectors $\vec{a}_{1,2,3}$ read
\begin{equation}
\begin{pmatrix}
\vec{a}_1 \\
\vec{a}_2 \\
\vec{a}_3
\end{pmatrix} =  \frac{a}{2}
\begin{pmatrix}
0 & 1 & 1\\
1 & 0 & 1\\
1 & 1 & 0
\end{pmatrix} \begin{pmatrix}
\hat{x} \\
\hat{y} \\
\hat{z}
\end{pmatrix},
\end{equation}
where $\hat{x},\hat{y},\hat{z}$ are the Cartesian vectors and $a$ is the lattice constant. 
The reciprocal lattice vectors $\vec{b_1}, \vec{b_2}, \vec{b_3}$ are expressed as \hspace{0.1cm}$\vec{b_1} = \frac{2\pi}{a}(\hat{x} - \hat{y} + \hat{z})$, \hspace{0.1cm}$\vec{b_2} = \frac{2\pi}{a}(\hat{x} + \hat{y} - \hat{z})$, \hspace{0.1cm}$\vec{b_3} = \frac{2\pi}{a}(-\hat{x} + \hat{y} + \hat{z})$.

\begin{table}[b]
\begin{tabular*}{0.48\textwidth}{@{\extracolsep{\fill}}lcc}
\hline \hline
 [eV] & \   LiBiZn \  & \  LiSbZn \  \\
\hline
$\varepsilon_{s}$  & -1.9 & -1.4 \\
$\varepsilon_{p}$  & -2.1 & -2.2\\
$\nu_{sp}$  & 0.9 & -0.9 \\
$\lambda_{soc}$  & -0.4 & -0.1 \\
$t_{s}$ & 0.1 &  0.2 \\
$t^{(d)}_{p}$  & 0.1 & 0.1 \\
$t^{(od)}_{p}$  & 0.1 & 0.1 \\
$t_{sp}$  & 0.9 & -0.9\\
\hline \hline
\end{tabular*}
\caption{Tight-binding parameters for LiBiZn and LiSbZn.}\label{tab:params}
\end{table}

The resulting Hamiltonian involves diagonal energies for $s$ ($\varepsilon_s$) and $p$ orbitals ($\varepsilon_p$), a local spin-orbit coupling $\lambda_{soc}$ in the $p$ shell, hoppings between $s$ orbitals ($t_s$), between identical ($t^{(d)}_p$) or different ($t^{(od)}_p$)  $p$ orbitals, and between $s$ and $p$ orbitals ($t_{sp}$) as well as a local hybridization between $s$ and $p$ orbitals ($\nu_{sp}$).
The values adopted by these parameters in LiBiZn and LiSbZn are presented in Table \ref{tab:params}.
The Hamiltonian in momentum space reads
\begin{equation}\label{eq:TBham}
\begin{split}
	&\mathcal{H}(\mathbf{k}) = \mathcal{H}_{0}+ \\ 
	&  \big(t_{sp} h_{sp,001} + t^{(od)}_{p} h^{(od)}_{p,001} + t_{s} \mathbf{1}_s + t^{(d)}_{p} h^{(d)}_{p,001}\big) e^{i \frac{(k_x + k_y)}{2}} + \\
	& \big(t_{sp} h_{sp,010} + t^{(od)}_{p} h_{p,010}^{(od)} + t_{s}  \mathbf{1}_s + t^{(d)}_{p} h_{p,010}^{(d)}\big) e^{i \frac{(k_x + k_z)}{2}} + \\
	& \big(t_{sp} h_{sp,100} + t^{(od)}_{p} h^{(od)}_{p,100} + t_{s}  \mathbf{1}_s + t^{(d)}_{p} h^{(d)}_{p,100}\big) e^{i \frac{(k_y + k_z)}{2}} +\\
	&  \big(t^{(od)}_{p} h^{(od)}_{p,1-10} + t_{s} \mathbf{1}_{s} + t^{(d)}_{p} h^{(d)}_{p,001}\big) e^{i \frac{(k_y - k_x)}{2}}+ \\
	&  \big(t^{(od)}_{p} h^{(od)}_{p,10-1} + t_{s} \mathbf{1}_{s} + t^{(d)}_{p}  h_{p,010}^{(d)}\big) e^{i \frac{(k_z - k_x)}{2}} + \\
	& \big(t^{(od)}_{p}  h^{(od)}_{p,01-1} + t_{s} \mathbf{1}_{s} + t^{(d)}_{p} h^{(d)}_{p,100}\big) e^{i \frac{(k_z - k_y)}{2}} + \\
& + h.c.
\end{split}
\end{equation}
where
\begin{equation}
	\mathcal{H}_{0} =  \varepsilon_{s} \mathbf{1}_s + \varepsilon_{p} \mathbf{1}_p +  \nu_{sp} V_{sp} + \lambda_{soc} \Gamma
\end{equation}
and matrices introduced are
\[
	\mathbf{1}_s=\begin{bmatrix}
 \mathbf{1}_{2\times 2} & \mathbf{0}_{6\times 6}\\
 \mathbf{0}_{6\times 6}& \mathbf{0}_{6\times 6} \\
\end{bmatrix}
\]
\[
	\mathbf{1}_p=\begin{bmatrix}
 \mathbf{0}_{2\times 2} & \mathbf{0}_{6\times 6}\\
 \mathbf{0}_{6\times 6}& \mathbf{1}_{6\times 6} \\
\end{bmatrix}
\]
\begin{equation}
	V_{sp} =
\begin{pmatrix}
	0 & 0 & 1 & 0 & 1 & 0 & 1 & 0\\
	0 &  0 & 0 & 1 & 0 & 1 & 0 & 1\\
	1 & 0 & 0 & 0 & 0 & 0 & 0 & 0\\
	0 & 1 & 0 & 0 & 0 & 0 & 0 & 0\\
	1 & 0 & 0 & 0 & 0 & 0 & 0 & 0 \\
	0 & 1 & 0 & 0 & 0 & 0 & 0 & 0 \\
	1 & 0 & 0 & 0 & 0 & 0 & 0 & 0 \\
	0 & 1 & 0 & 0 & 0 & 0 & 0 & 0 \\
\end{pmatrix}
\end{equation}
\begin{equation} 
\Gamma =
\begin{pmatrix}
	0 & 0 & 0 & 0 & 0 & 0 & 0 & 0\\
	0 &  0 & 0 & 0 & 0 & 0 & 0 & 0\\
	0 & 0 & 0 & 0 & 0 & i & -i & 0\\
	0 & 0 & 0 & 0 & i & 0 & 0 & i\\
	0 & 0 & 0 & -i & 0 & 0 & 0 & 1 \\
	0 & 0 & -i & 0 & 0 & 0 & -1 & 0 \\
	0 & 0 & i & 0 & 0 & -1 & 0 & 0 \\
	0 & 0 & 0 & -i & 1 & 0 & 0 & 0 \\
\end{pmatrix}
\end{equation}
\begin{equation}
h_{sp,001} =
\begin{pmatrix}
	0 & 0 & 0 & 0 & 0 & 0 & 0 & 0\\
	0 &  0 & 0 & 0 & 0 & 0 & 0 & 0\\
	1 & 0 & 0 & 0 & 0 & 0 & 0 & 0\\
	0 & 1 & 0 & 0 & 0 & 0 & 0 & 0\\
	-1 & 0 & 0 & 0 & 0 & 0 & 0 & 0 \\
	0 & -1 & 0 & 0 & 0 & 0 & 0 & 0 \\
	1 & 0 & 0 & 0 & 0 & 0 & 0 & 0 \\
	0 & 1 & 0 & 0 & 0 & 0 & 0 & 0 \\
\end{pmatrix}
\end{equation}
\begin{equation}
	h^{(od)}_{p,001} =
\begin{pmatrix}
	0 & 0 & 0 & 0 & 0 & 0 & 0 & 0\\
	0 &  0 & 0 & 0 & 0 & 0 & 0 & 0\\
	0 & 0 & 0 & 0 & -1 & 0 & 2 & 0\\
	0 & 0 & 0 & 0 & 0 & -1 & 0 & 2\\
	0 & 0 & 1 & 0 & 0 & 0 & 1 & 0 \\
	0 & 0 & 0 & 1 & 0 & 0 & 0 & 1 \\
	0 & 0 & 2 & 0 & -1 & 0 & 0 & 0 \\
	0 & 0 & 0 & 2 & 0 & -1 & 0 & 0 \\
\end{pmatrix}
\end{equation}
\begin{equation}
	h^{(d)}_{p,001} =
\begin{pmatrix}
	0 & 0 & 0 & 0 & 0 & 0 & 0 & 0\\
	0 &  0 & 0 & 0 & 0 & 0 & 0 & 0\\
	0 & 0 & 1 & 0 & 0 & 0 & 0 & 0\\
	0 & 0 & 0 & 1 & 0 & 0 & 0 & 0\\
	0 & 0 & 0 & 0 & 2 & 0 & 0 & 0 \\
	0 & 0 & 0 & 0 & 0 & 2 & 0 & 0 \\
	0 & 0 & 0 & 0 & 0 & 0 & 1 & 0 \\
	0 & 0 & 0 & 0 & 0 & 0 & 0 & 1 \\
\end{pmatrix}
\end{equation}
\begin{equation}
	h^{(od)}_{p,1-10} =
\begin{pmatrix}
	0 & 0 & 0 & 0 & 0 & 0 & 0 & 0\\
	0 &  0 & 0 & 0 & 0 & 0 & 0 & 0\\
	0 & 0 & 0 & 0 & 1 & 0 & -2 & 0\\
	0 & 0 & 0 & 0 & 0 & 1 & 0 & -2\\
	0 & 0 & -1 & 0 & 0 & 0 & 1 & 0 \\
	0 & 0 & 0 & -1 & 0 & 0 & 0 & 1 \\
	0 & 0 & -2 & 0 & -1 & 0 & 0 & 0 \\
	0 & 0 & 0 & -2 & 0 & -1 & 0 & 0 \\
\end{pmatrix}
\end{equation}
Above, we have only shown the terms which arise from independent tight-binding parameters.
The other hopping matrices can be connected by the symmetries of the Hamiltonian. 
For instance, $C_{3,111} h_{sp,001} C_{3,111}^{-1} = h_{sp,100}, C_{3,111} h^{(od)}_{p,001} C_{3,111}^{-1} = h^{(od)}_{p,100}, C_{3,111} h^{(od)}_{p,1-10} C_{3,111}^{-1}= h^{(od)}_{p,01-1}$, and so on. 

We now explain the symmetries of this Hamiltonian.
In the cubic phase, the system has two-fold rotational symmetries with respect to the cubic edges (e.g., $C_{2z}$), three-fold rotational symmetries with respect to the cubic diagonals ($C_{3,111}$), reflection symmetries ($M_{xy}$), fourfold rotoinversion symmetries ($C_{4z}I$), and time-reversal symmetry ($\mathcal{T}$). 
In cartesian form, the Hamiltonian $\mathcal{H}(k_x, k_y, k_z)$ satisfies:
\begin{equation} 
C_{2z} \mathcal{H}(k_x, k_y, k_z) C_{2z}^{-1} = \mathcal{H}(-k_x, -k_y, k_z) 
\end{equation}
\begin{equation} 
C_{3,111} \mathcal{H}(k_x, k_y, k_z) C_{3, 111}^{-1}  = \mathcal{H}(k_z, k_x, k_y) 
\end{equation}
\begin{equation} 
M_{xy} \mathcal{H}(k_x, k_y, k_z) M_{xy}^{-1} = \mathcal{H}(-k_y, -k_x, k_z)
\end{equation}
\begin{equation} 
(C_{4z}I) \mathcal{H}(k_x, k_y, k_z)  (C_{4z} I)^{-1} = \mathcal{H}(k_y, -k_x, -k_z) 
\end{equation}
\begin{equation}
\mathcal{T} \mathcal{H}(k_x, k_y, k_z) \mathcal{T}^{-1} = \mathcal{H}(-k_x,- k_y,- k_z)
\end{equation}
The above symmetry constraints translate to the following set of equations, if the momenta are  written in the basis of the reciprocal lattice vectors, coordinates which we name ($k_1$, $k_2$, $k_3$).
\begin{equation}
C_{2z} \mathcal{H}(k_1, k_2, k_3) C_{2z}^{-1} = \mathcal{H}(k_3, -k_1-k_2-k_3, k_1)
\end{equation}
\begin{equation}
C_{3,111} \mathcal{H}(k_1, k_2, k_3) C_{3,111}^{-1} = \mathcal{H}(k_3, k_1, k_2)
\end{equation}
\begin{equation}
M_{xy} \mathcal{H}(k_1, k_2, k_3) M_{xy}^{-1} = \mathcal{H}(k_1, -k_1-k_2-k_3, k_3)
\end{equation}
\begin{equation}
(C_{4z}I) \mathcal{H}(k_1, k_2, k_3)  (C_{4z} I)^{-1} = \mathcal{H}(-k_1-k_2-k_3, k_1, k_2) 
\end{equation}
\begin{equation}
\mathcal{T} \mathcal{H}(k_1, k_2, k_3) \mathcal{T}^{-1} = \mathcal{H}(-k_1,- k_2,- k_3),
\end{equation}
where we have used the relations;\hspace{0.1cm}$\hat{k}_{x}=\frac{a}{2\pi}(\frac{\vec{k_1}+\vec{k_2}}{2})$,\hspace{0.1cm}$\hat{k}_{y}=\frac{a}{2\pi}(\frac{\vec{k_2}+\vec{k_3}}{2})$,\hspace{0.2cm}$\hat{k}_{z}=\frac{a}{2\pi}(\frac{\vec{k_3}+\vec{k_1}}{2})$.
Finally, the representation used for the time-reversal operator in the 8-dimensional Hilbert space is
 \begin{equation} 
\mathcal{T}=
\begin{pmatrix}
	0 & -1 & 0 & 0 & 0 & 0 & 0 & 0\\
	1 &  0 & 0 & 0 & 0 & 0 & 0 & 0\\
	0 & 0 & 0 & -1 & 0 & 0 & 0 & 0\\
	0 & 0 & 1 & 0 & 0 & 0 & 0 & 0\\
	0 & 0 & 0 & 0 & 0 & -1 & 0 & 0 \\
	0 & 0 & 0 & 0 & 1 & 0 & 0 & 0 \\
	0 & 0 & 0 & 0 & 0 & 0 & 0 & -1 \\
	0 & 0 & 0 & 0 & 0 & 0 & 1 & 0 \\ 
\end{pmatrix} \mathcal{K}
\end{equation}
where $\mathcal{K}$ is the complex conjugation operator, while the representation of the unitary symmetries are
\begin{equation} 
C_{3,111} = \frac{1}{2}
\begin{pmatrix}
	\alpha & -\alpha^* & 0 & 0 & 0 & 0 & 0 & 0\\
	\alpha & \alpha^*  & 0 & 0 & 0 & 0 & 0 & 0\\
	0 & 0 & 0 & 0 & 0 & 0 & \alpha &  -\alpha^*\\
	0 & 0 & 0 & 0 & 0 & 0 &  \alpha & \alpha^*\\
	0 & 0 & \alpha & -\alpha^* & 0 & 0 & 0 & 0 \\
	0 & 0 &  \alpha & \alpha^* & 0 & 0 & 0 & 0 \\
	0 & 0 & 0 & 0 & \alpha & -\alpha^* & 0 & 0 \\
	0 & 0 & 0 & 0 & \alpha & \alpha^* & 0 & 0 \\
\end{pmatrix}
\end{equation}
with $\alpha=1-i$,
\begin{align}
	& C_{4z}I = \frac{1}{\sqrt{2}}\times \nonumber \\ 
	& 
\begin{pmatrix}
	\alpha^* & 0 & 0 & 0 & 0 & 0 & 0 & 0\\
	0 &  \alpha & 0 & 0 & 0 & 0 & 0 & 0\\
	0 & 0 & 0 & 0 & 0 & 0 & \alpha^*\beta_{yz} & 0\\
	0 & 0 & 0 & 0 & 0 & 0 & 0 & \alpha \beta_{yz}\\
	0 & 0 & 0 & 0 & -\alpha^*\beta_{yz} & 0 & 0 & 0 \\
	0 & 0 & 0 & 0 & 0 & -\alpha\beta_{yz} & 0 & 0 \\
	0 & 0 & -\alpha^*\beta_{yz} & 0 & 0 & 0 & 0 & 0 \\
	0 & 0 & 0 & -\alpha \beta_{yz} & 0 & 0 & 0 & 0 \\
\end{pmatrix}
\end{align}
where	$\beta_{yz}=e^{-i(k_y+k_z)}$,
\begin{equation} 
C_{2z} = 
\begin{pmatrix}
	i & 0 & 0 & 0 & 0 & 0 & 0 & 0\\
	0 & -i & 0 & 0 & 0 & 0 & 0 & 0\\
	0 & 0 & -i\beta_{xy} & 0 & 0 & 0 & 0 & 0\\
	0 & 0 & 0 & i\beta_{xy} & 0 & 0 & 0 & 0\\
	0 & 0 & 0 & 0 & i\beta_{xy} & 0 & 0 & 0 \\
	0 & 0 & 0 & 0 & 0 & -i\beta_{xy} & 0 & 0 \\
	0 & 0 & 0 & 0 & 0 & 0 & -i\beta_{xy} & 0 \\
	0 & 0 & 0 & 0 & 0 & 0 & 0 & i\beta_{xy} \\
\end{pmatrix}
\end{equation}
where $\beta_{xy}=e^{-i(k_x+k_y)}$, and
\begin{align} 
	& M_{xy} = \frac{1}{\sqrt{2}}\times\nonumber \\
	&\begin{pmatrix}
	0 & -\alpha^* & 0 & 0 & 0 & 0 & 0 & 0\\
	\alpha &  0 & 0 & 0 & 0 & 0 & 0 & 0\\
	0 & 0 & 0 & 0 & 0 & 0 & 0 & \alpha^*\beta_{xy}\\
	0 & 0 & 0 & 0 & 0 & 0 & -\alpha\beta_{xy} & 0\\
	0 & 0 & 0 & 0 & 0 & -\alpha^*\beta_{xy} & 0 & 0 \\
	0 & 0 & 0 & 0 & \alpha\beta_{xy} & 0 & 0 & 0 \\
	0 & 0 & 0 & \alpha^*\beta_{xy} & 0 & 0 & 0 & 0 \\
	0 & 0 & -\alpha\beta_{xy} & 0 & 0 & 0 & 0 & 0 \\
\end{pmatrix}
\end{align}.

\section{Momentum-resolved weight functions in LiBiZn}
\label{app:weights}

\begin{figure}[tb]
\centering
\includegraphics[width=0.8\columnwidth]{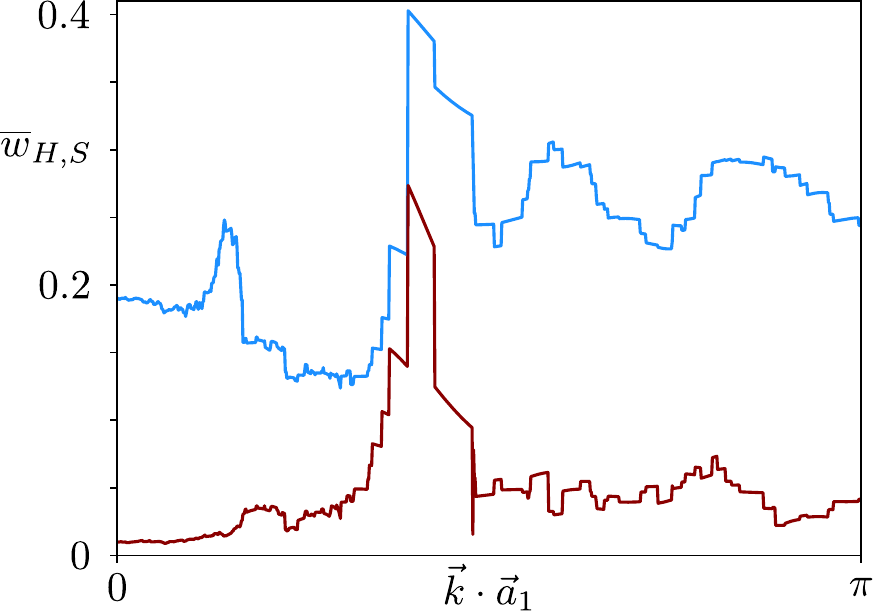}
\caption{
The LiBiZn weight functions $\overline{w}_{\text{H}}$ and $\overline{w}_{\text{S}}$ are averaged over all states with energies $|\varepsilon| < 0.05$ eV, in the case of the prism geometry of Fig.~\ref{fig:rhombus}. 
The dark red color denotes $\overline{w}_{\text{H}}$, whereas the light blue color represents $\overline{w}_{\text{S}}$.
The vertical jumps are finite-size effects.
They correspond to momenta at which the number of states in the energy interval changes.
}
\label{fig:weightsLiBiZn} 
\end{figure}

Here we present additional data on the momentum-resolved, averaged weight functions $\overline{w}_{\text{H}}$ and $\overline{w}_{\text{S}}$, in the case of LiBiZn in the prism geometry of Fig.~\ref{fig:rhombus}.
The results in Fig.~\ref{fig:weightsLiBiZn} show the two weights, averaged over all states in the energy interval $[-0.05, 0.05]$ eV, as a function of $\vec{k}\cdot\vec{a}_1$.
While close to the Brillouin zone there is almost no weight on the hinge sites, we observe a peak in $\overline{w}_{\text{H}}$ around $\vec{k}\cdot\vec{a}_1 = 1.25$, corresponding to the light-colored regions close to $\varepsilon=0$ in the bandstructure of Fig.~\ref{fig:rhombus}(b).
At this peak, $\overline{w}_{\text{H}} > 0.27$ and $\overline{w}_{\text{S}} > 0.4$, indicating that more than two thirds of the probability density is concentrated on the boundary of the prism, with more than one quarter concentrated at the hinge sites.

\section{Alternate prism geometry}
\label{app:rectangle}

\begin{figure*}[t]
\centering
\includegraphics[width=\textwidth]{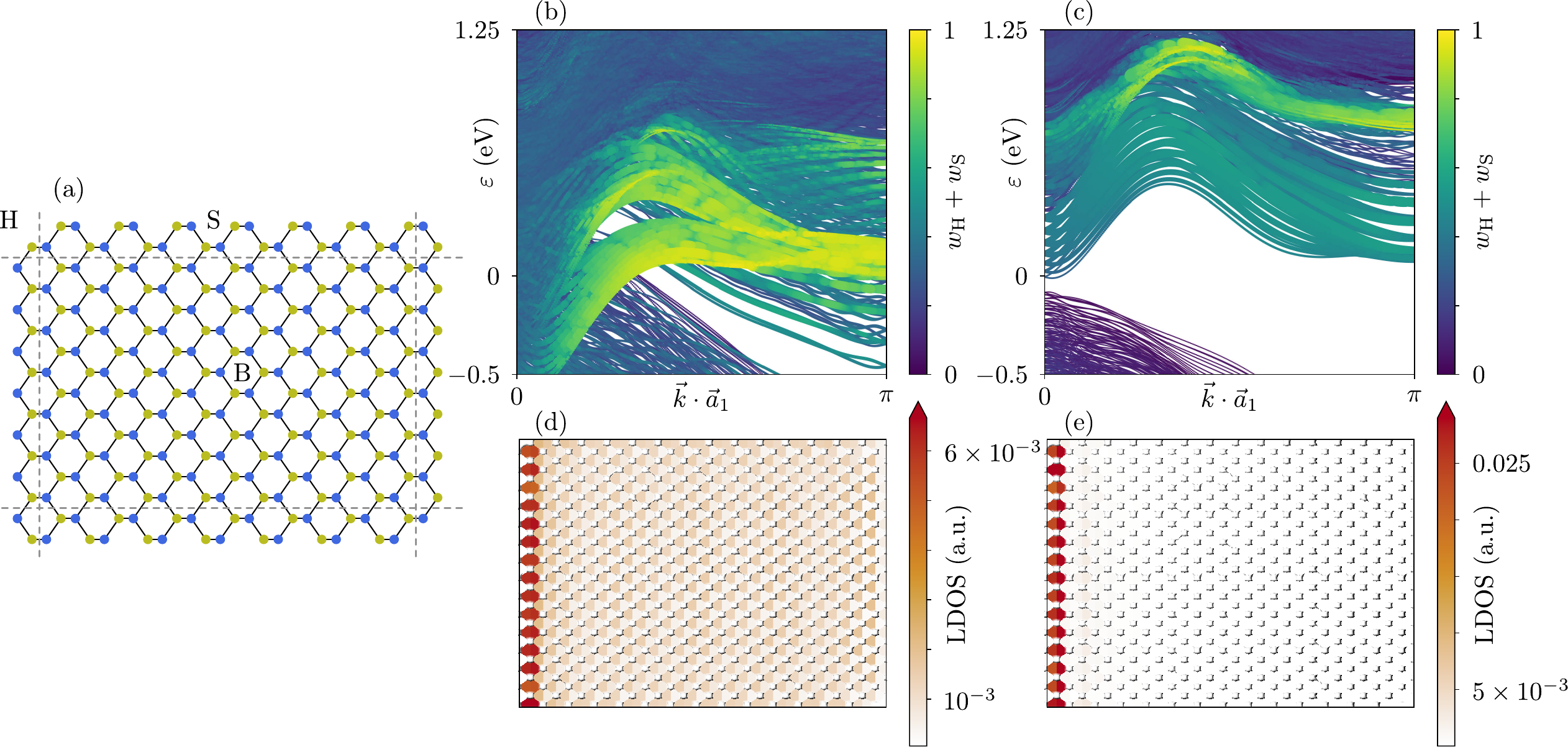}
\caption{
Same as Fig.~\ref{fig:rhombus}, but for a different geometry, as shown in panel (a).
Panels (b, c): Band structure in the absence of strain in the case of LiBiZn (b) and LiSbZn (c).
Panels (d, e): The local density of states of the infinite prism, averaged over momentum, is shown for LiBiZn at $E=0$ (d), and for LiSbZn at $E=0.2$ eV (e).}
\label{fig:rectangle} 
\end{figure*}

Here we show the results for the band structure and LDOS in the case of an infinite prism with a rectangular cross-section (Fig.~\ref{fig:rectangle}). 
We find that the hinge modes are now spread over the entire Bi(Sb)-terminated surface.

\end{document}